\documentclass[10pt,conference]{IEEEtran}
\IEEEoverridecommandlockouts

\usepackage{cite}
\usepackage[T1]{fontenc}
\usepackage{amsmath,amssymb,amsfonts}
\usepackage{graphicx}
\usepackage{textcomp}
\usepackage{xcolor}
\def\BibTeX{{\rm B\kern-.05em{\sc i\kern-.025em b}\kern-.08em
    T\kern-.1667em\lower.7ex\hbox{E}\kern-.125emX}}

\usepackage[caption=false,font=footnotesize]{subfig}
\usepackage[bookmarks=false]{hyperref}
\usepackage{tikz}
\usetikzlibrary{shapes,arrows}
\usetikzlibrary{positioning,calc}
\usepackage{pgfplots}
\pgfplotsset{compat=1.18}
\usepackage{booktabs}
\usepackage{url}
\usepackage{orcidlink}
\usepackage[table]{xcolor}
\usepackage{multirow}
\usepackage{threeparttable}

\usepackage[linesnumbered, ruled, vlined]{algorithm2e}

\begin{document}

\title{Orthogonal Code-Based Parallel Signaling for PLCA in Shared-Medium Ethernet}

\author{
  \IEEEauthorblockN{Mohamed Elattar\IEEEauthorrefmark{1}\IEEEauthorrefmark{2}, Ahmed Abdelkader\IEEEauthorrefmark{1}, Zhe Lou\IEEEauthorrefmark{1}, Georg Carle\IEEEauthorrefmark{2}}
  
  \IEEEauthorblockA{\IEEEauthorrefmark{1}Huawei Heisenberg Research Center, Munich, Germany}
  
  \IEEEauthorblockA{\IEEEauthorrefmark{2} Technical University of Munich, Munich, Germany}
    
    \IEEEauthorblockA{\IEEEauthorrefmark{1}\{mohamed.osama.elattar, ahmed.abdelkader3, zhe.lou\}@huawei.com, \IEEEauthorrefmark{2}m.elattar@tum.de, \IEEEauthorrefmark{2}carle@net.in.tum.de }

\thanks{This work has been submitted to the IEEE for possible publication. Copyright may be transferred without notice, after which this version may no longer be accessible.}

}

\maketitle
\begin{abstract}
Physical Layer Collision Avoidance (PLCA) enables collision-free channel access in shared-medium Ethernet by granting each node an exclusive transmission opportunity (TO) in a sequential, cyclic manner. This sequential, node-based access structure, however, is the root cause of PLCA's fundamental limitations, causing access latency to grow with network size and idle TOs to accumulate when traffic is sparse, a limitation inherited by existing PLCA extensions. To address this, we propose OC-PLCA, an orthogonal code-based parallel signaling mechanism for PLCA. Rather than signaling sequentially, all nodes with pending traffic transmit their unique Walsh-Hadamard code simultaneously to signal their intent to transmit. Since these codes are mutually orthogonal, each node independently identifies the complete set of active nodes, allowing the transmission cycle to include only active nodes and eliminating idle TOs entirely. We further investigate the impact of quantization on active node detection, and propose a grouping-based extension that preserves unambiguous detection for networks of an arbitrary number of nodes. Theoretical analysis shows that OC-PLCA reduces signaling overhead by 88\% for an 8-node network, with the reduction growing as network size increases. Simulation results confirm significant reductions in observed access latency across a range of network loads and heterogeneous node activity.
\end{abstract}


\section{Introduction}
\label{Section: Introduction}

Shared-medium Ethernet is increasingly adopted across automotive and industrial environments, as well as emerging domains such as humanoid robotics. In automotive networks, the shift towards zonal architectures has made Ethernet a dominant in-vehicle communication technology, progressively replacing legacy technologies such as the Controller Area Network (CAN)\cite{automotive-architectures}. A crucial enabler of this transition is the standardization of Physical Layer Collision Avoidance (PLCA) in 10BASE-T1S \cite{8023cg, 8023da}. PLCA grants each node on the shared channel an exclusive transmission opportunity (TO) in a sequential, cyclic manner. During its TO, a node signals its intent to transmit; otherwise, the cycle advances to the next node. By ensuring that at most one node accesses the channel at any given time, PLCA eliminates collisions and bounds access latency, making it attractive for latency-sensitive and safety-critical applications.

This sequential, node-based access structure, however, is the root cause of PLCA's fundamental limitations, resulting in suboptimal access performance under both light and heavy load conditions. Under heavy load, the worst-case access latency grows directly with the number of nodes on the network, limiting scalability in large deployments. Under light load, the cycle nonetheless advances through every device regardless of whether it has pending traffic, introducing substantial idle time on the channel that likewise grows with network size.

Several extensions have been proposed to introduce prioritization into PLCA \cite{PLCAMIDs, PLCANP, PriorityBasedPLCA, weighted-Multi-ID}. However, these mechanisms retain the same sequential, node-based access structure as standard PLCA, and therefore inherit its long access latency. Moreover, to support prioritization, such mechanisms typically introduce additional signaling, further increasing overhead and reducing channel utilization.

To address these limitations, this paper proposes OC-PLCA, an orthogonal code-based parallel signaling mechanism for PLCA. Rather than signaling sequentially, each node is assigned a unique orthogonal code, such as a Walsh-Hadamard (WH) code, and all nodes with pending traffic signal their intent to transmit simultaneously at the beginning of each cycle. Since the codes are mutually orthogonal, simultaneous transmissions do not interfere, allowing the set of active nodes for the current cycle to be identified through simple correlation checks at each node. Following this signaling phase, only the identified active nodes transmit sequentially, thereby eliminating idle TOs from the cycle entirely. 

The remainder of this paper is structured as follows. Section~\ref{Section: Background} reviews related work. Section~\ref{Section: System Model} presents the system model, including PLCA and WH codes. Section~\ref{Section: Novel} introduces the proposed mechanism and investigates the quantization impact on active node detection.  Evaluation is presented in Section~\ref{Section: Evaluation}. Finally, Section~\ref{Section: Conclusion} concludes the paper.

\section{Background} 
\label{Section: Background}

Channel access in shared-medium Ethernet has shifted from stochastic contention-based access through Carrier Sense Multiple Access with Collision Detection (CSMA/CD) to deterministic, collision-free access via PLCA. PLCA allocates an exclusive TO to each node in a cyclic order determined by its node identifier (ID), ensuring bounded-latency access. However, PLCA's sequential, node-based structure introduces fundamental limitations. First, the worst-case access latency grows directly with the number of devices, limiting scalability in large networks. Second, the transmission order is independent of traffic priority; therefore, latency-sensitive traffic may be delayed by lower-priority transmissions. Lastly, because the cycle traverses every node regardless of whether it has pending data, idle TOs accumulate when traffic is sparse, introducing substantial idle time on the channel.  

Several extensions have been proposed to introduce priority awareness into PLCA and reduce access latency for high-priority nodes. Some allow designated priority nodes to interrupt the ongoing cycle, either during the Inter-Packet Gap \cite{PLCANP, priority_plca} or by preempting an ongoing transmission \cite{InterruptPLCA}, to gain earlier medium access. Others reorder access by assigning node IDs according to device priority, allowing high-priority devices to restart the cycle ahead of others \cite{PriorityBasedPLCA}, or increase a device's transmission frequency by assigning it multiple node IDs \cite{PLCAMIDs, weighted-Multi-ID}. Another approach introduces a signaling phase in which devices advertise the priority of their pending traffic, allowing TOs to be reordered within the cycle accordingly~\cite{PriorityAdvertisingPLCA}. Other work integrates advanced scheduling mechanisms into PLCA, such as time-aware shaping~\cite{TASPLCA}, strict priority scheduling~\cite{PLCA_SP}, credit-based shaping~\cite{CBS-PLCA}, and per-flow shaping~\cite{Flow-shaper-PLCA}.

Despite their differences, all of these mechanisms retain the same sequential, node-based access structure as standard PLCA, and therefore inherit its long access-waiting times. Moreover, most of them introduce additional signaling to support prioritization, further increasing overhead and reducing channel utilization.

\section{System Model} 
\label{Section: System Model}

This section presents the system model, including the network model, the PLCA mechanism, and the generation and orthogonality properties of WH codes.

\subsection{Network Model}
\label{Subsection: Network and Traffic Model}

We consider a shared-medium (multidrop) Ethernet network consisting of a set of nodes $\mathcal{N}$, where $N=|\mathcal{N}|$, connected to a common channel with rate $R$ and sample duration $T_s = 1/R$. All nodes share a single collision domain, such that at most one node may transmit at any given time to ensure successful reception. Each node is assigned a unique ID $n \in \{0,\dots,N-1 \}$, which determines its order in the standard PLCA access cycle. 

We assume that all nodes are synchronized and that propagation delays are negligible, such that simultaneously transmitted codewords arrive at any destination with a relative delay of less than half a sample duration, combine additively, and have approximately equal amplitude, preserving their mutual orthogonality. Such effects are not addressed in this paper and are left for future investigation.

\subsection{PLCA: Physical Layer Collision Avoidance}
\label{Subsection: PLCA}

Each node maintains a local TO counter $\hat{n}$ that tracks the currently active TO. At the beginning of each cycle, the coordinator (typically the node with $n=0$) broadcasts a BEACON message of duration~$T_\mathrm{b}$, synchronizing all nodes and resetting their TO counters to $\hat{n}=0$. The counter then increments sequentially across all nodes. When the counter matches a node's ID ($\hat{n}=n$), that node is granted exclusive access to the medium. If the node has pending data, it transmits a COMMIT message of duration $T_\mathrm{c}$ to reserve the channel, followed by the data packet. Otherwise, it remains silent for a predefined idle interval $T_\mathrm{i}$, yielding its TO. After either a data transmission or the idle interval, the TO counter advances to the next node. Once the last TO ($\hat{n}=N-1$) completes, a new BEACON message is issued to initiate the next cycle, as illustrated in Fig.~\ref{fig: Operation and transmission sequence of PLCA}.

The sequential, node-based access structure of PLCA introduces substantial overhead, characterized by two distinct worst cases that arise under opposite load conditions. Under heavy load, the worst-case access latency\footnote{Access latency is defined as the time elapsed between the instant the head-of-line packet at a node becomes ready for transmission and the instant its transmission begins.} arises when all $N-1$ preceding nodes have pending data and transmit packets of the maximum size (Fig.~\ref{subfig: Worst-case access latency in PLCA}), which can be expressed as:
\begin{equation}
\label{equ: PLCA maximum latency}
L_{n,\max}^{\mathrm{PLCA}} = T_\mathrm{b} + N T_\mathrm{c} + (N-1) T_{\mathrm{pkt,max}},
\end{equation}
where $T_{\mathrm{pkt,max}}$ denotes the transmission time of the maximum-sized packet. Under light load, the worst-case signaling overhead\footnote{Signaling overhead is defined as the cumulative duration, within the waiting time, attributable exclusively to non-data transmission elements, including BEACON and COMMIT messages as well as idle intervals.} arises when all $N-1$ preceding nodes have no pending data and yield their TOs (Fig.~\ref{subfig: Transmission sequence of standard PLCA}), resulting in:
\begin{equation}
\label{equ: PLCA maximum overhead}
O_{n,\max}^{\mathrm{PLCA}} = T_\mathrm{b} + (N-1) T_\mathrm{i} + T_\mathrm{c}.
\end{equation}

As evident from (\ref{equ: PLCA maximum latency}) and (\ref{equ: PLCA maximum overhead}), both the worst-case access latency and the worst-case signaling overhead grow linearly with~$N$, confirming that the sequential access structure of PLCA results in suboptimal access latency and channel utilization regardless of network load.

\subsection{Walsh-Hadamard Codes}
\label{Subsection: Walsh-Hadamard Codes}

A Hadamard matrix $\mathbf{H}_m$ of dimension $2^m \times 2^m$, where $m \in \mathbb{Z}_{\ge0}$, is constructed recursively from the seed $\mathbf{H}_0$ as:
\begin{equation}
    \mathbf{H}_0=[1], \quad
    \mathbf{H}_m = 
    \begin{bmatrix} 
        \mathbf{H}_{m-1} & \mathbf{H}_{m-1} \\ 
        \mathbf{H}_{m-1} & -\mathbf{H}_{m-1} 
    \end{bmatrix}, \quad m \geq 1.
\end{equation}

Each column of $\mathbf{H}_m$ constitutes a WH codeword of length~$2^m$, giving a codebook $\mathcal{C}$ of $2^m$ codewords, denoted $\mathbf{c}_i\in\{+1,-1\}^{2^m}$. WH codes exhibit perfect orthogonality. The inner product between any two codewords $\mathbf{c}_i,\mathbf{c}_j \in \mathcal{C}$ satisfies:
\begin{equation}
\mathbf{c}_i^T  \mathbf{c}_j = \sum_{k=0}^{2^m-1} c_{i,k} \cdot c_{j,k} = \begin{cases} 2^m, & i = j, \\ 0, & i \neq j, \end{cases}
\end{equation}
where $c_{i,k}$ denotes the $k$-th element of the codeword $\mathbf{c}_i$. This zero cross-correlation ensures that simultaneously transmitted codewords produce no mutual interference and can be perfectly separated at the receiver through correlation-based detection.

Consider a scenario in which a subset $\mathcal{A}\subseteq \mathcal{N}$ of nodes simultaneously transmits their assigned codewords over an ideal shared channel, modeled as a simple channel gain $h$ and additive white Gaussian noise. The sampled received signal, given by the column vector $\mathbf{r}$ of length $2^m$, can be written as:
\begin{equation}
\mathbf{r} = h \sum_{n \in \mathcal{A}}  \mathbf{c}_n + \mathbf{w},
\label{equ: received signal r}
\end{equation}

where $\mathbf{w}$ denotes the sampled noise column vector of length~$2^m$. The receiver identifies active nodes by correlating~$\mathbf{r}$ with each codeword:
\begin{equation}
y_i = \mathbf{r}^T\mathbf{c}_i = h\sum_{n \in \mathcal{A}} \mathbf{c}_n^T \mathbf{c}_i + \mathbf{w}^T \mathbf{c}_i.
\end{equation}
By orthogonality, and assuming a unity channel gain ($h=1$) and negligible noise for simplicity, the correlator output is given by:
\begin{equation}
y_i = \begin{cases} 2^m, & i \in \mathcal{A}, \\ 0, & i \notin \mathcal{A}, \end{cases}
\end{equation}
allowing the receiver to perfectly identify the set of active nodes.

\definecolor{GrayBlueOLL}{HTML}{44546A}
\definecolor{GrayBlueOLM}{HTML}{333f50}
\definecolor{GrayBlueOLD}{HTML}{222A35}

\definecolor{GrayBlueFillL}{HTML}{D6DCE5}
\definecolor{GrayBlueFillD}{HTML}{ADB9CA}

\definecolor{BlueOLL}{HTML}{5B9BD5}
\definecolor{BlueOLM}{HTML}{2E75B6}
\definecolor{BlueOLD}{HTML}{1F4E79}

\definecolor{BlueFillL}{HTML}{DEEBF7}
\definecolor{BlueFillD}{HTML}{BDD7EE}

\definecolor{GreenOLL}{HTML}{70AD47}
\definecolor{GreenOLM}{HTML}{548235}
\definecolor{GreenOLD}{HTML}{385723}

\definecolor{GreenFillL}{HTML}{E2F0D9}
\definecolor{GreenFillD}{HTML}{C5E0B4}

\definecolor{WhiteOLL}{HTML}{BFBFBF}
\definecolor{WhiteOLM}{HTML}{A6A6A6}
\definecolor{WhiteOLD}{HTML}{7F7F7F}

\definecolor{WhiteFillL}{HTML}{F2F2F2}
\definecolor{WhiteFillD}{HTML}{D9D9D9}

\definecolor{GrayOLL}{HTML}{A5A5A5}
\definecolor{GrayOLM}{HTML}{7C7C7C}
\definecolor{GrayOLD}{HTML}{525252}

\definecolor{GrayFillL}{HTML}{EDEDED}
\definecolor{GrayFillD}{HTML}{DBDBDB}

\definecolor{YellowOLL}{HTML}{FFC000}
\definecolor{YellowOLM}{HTML}{BF9000}
\definecolor{YellowOLD}{HTML}{7F6000}

\definecolor{YellowFillL}{HTML}{FFF2CC}
\definecolor{YellowFillD}{HTML}{FFE699}

\definecolor{OrangeOLL}{HTML}{ED7D31}
\definecolor{OrangeOLM}{HTML}{C55A11}
\definecolor{OrangeOLD}{HTML}{843C0C}

\definecolor{OrangeFillL}{HTML}{FBE5D6}
\definecolor{OrangeFillD}{HTML}{F8CBAD}

\definecolor{PriorityOL}{HTML}{C0392B} 
\definecolor{PriorityFill}{HTML}{F5B7B1}

\definecolor{CollisionOL}{HTML}{16A085} 
\definecolor{CollisionFill}{HTML}{A3E4D7}
\definecolor{GrayBlueOLL}{HTML}{44546A}
\definecolor{GrayBlueOLM}{HTML}{333f50}
\definecolor{GrayBlueOLD}{HTML}{222A35}

\definecolor{GrayBlueFillL}{HTML}{D6DCE5}
\definecolor{GrayBlueFillD}{HTML}{ADB9CA}

\definecolor{BlueOLL}{HTML}{5B9BD5}
\definecolor{BlueOLM}{HTML}{2E75B6}
\definecolor{BlueOLD}{HTML}{1F4E79}

\definecolor{BlueFillL}{HTML}{DEEBF7}
\definecolor{BlueFillD}{HTML}{BDD7EE}

\definecolor{GreenOLL}{HTML}{70AD47}
\definecolor{GreenOLM}{HTML}{548235}
\definecolor{GreenOLD}{HTML}{385723}

\definecolor{GreenFillL}{HTML}{E2F0D9}
\definecolor{GreenFillD}{HTML}{C5E0B4}

\definecolor{WhiteOLL}{HTML}{BFBFBF}
\definecolor{WhiteOLM}{HTML}{A6A6A6}
\definecolor{WhiteOLD}{HTML}{7F7F7F}

\definecolor{WhiteFillL}{HTML}{F2F2F2}
\definecolor{WhiteFillD}{HTML}{D9D9D9}

\definecolor{GrayOLL}{HTML}{A5A5A5}
\definecolor{GrayOLM}{HTML}{7C7C7C}
\definecolor{GrayOLD}{HTML}{525252}

\definecolor{GrayFillL}{HTML}{EDEDED}
\definecolor{GrayFillD}{HTML}{DBDBDB}

\definecolor{YellowOLL}{HTML}{FFC000}
\definecolor{YellowOLM}{HTML}{BF9000}
\definecolor{YellowOLD}{HTML}{7F6000}

\definecolor{YellowFillL}{HTML}{FFF2CC}
\definecolor{YellowFillD}{HTML}{FFE699}

\definecolor{OrangeOLL}{HTML}{ED7D31}
\definecolor{OrangeOLM}{HTML}{C55A11}
\definecolor{OrangeOLD}{HTML}{843C0C}

\definecolor{OrangeFillL}{HTML}{FBE5D6}
\definecolor{OrangeFillD}{HTML}{F8CBAD}

\definecolor{PriorityOL}{HTML}{C0392B} 
\definecolor{PriorityFill}{HTML}{F5B7B1}

\definecolor{CollisionOL}{HTML}{16A085} 
\definecolor{CollisionFill}{HTML}{A3E4D7}

\tikzstyle{beacon} = [rectangle, draw=YellowOLD, fill=YellowFillD, minimum height=0.6cm, minimum width=0.3cm]

\tikzstyle{commit} = [rectangle, draw=BlueOLD, fill=BlueFillD, minimum height=0.6cm, minimum width=0.1cm]

\tikzstyle{ipg_commit} = [rectangle, draw=BlueOLD, fill=BlueFillD, minimum height=0.5cm, minimum width=0.2cm]

\tikzstyle{CTO_commit} = [rectangle, draw=BlueOLD, fill=BlueFillD, minimum height=0.45cm, minimum width=0.1cm]

\tikzstyle{silent} = [rectangle, draw=WhiteOLD, fill=WhiteFillL, minimum height=0.6cm, minimum width=0.5cm]

\tikzstyle{ipg_silent} = [rectangle, draw=GrayOLD, fill=GrayFillD, minimum height=0.5cm, minimum width=0.3 cm]

\tikzstyle{CTO_silent} = [rectangle, draw=WhiteOLD, fill=WhiteFillL, minimum height=0.45cm, minimum width=0.3cm]

\tikzstyle{min_packet} = [rectangle, draw=GreenOLD, fill=GreenFillD, minimum   height=0.6cm, minimum width=0.8cm]

\tikzstyle{xsmall_packet} = [rectangle, draw=GreenOLD, fill=GreenFillD, minimum   height=0.6cm, minimum width=0.9cm]

\tikzstyle{small_packet} = [rectangle, draw=GreenOLD, fill=GreenFillD, minimum   height=0.6cm, minimum width=1cm]

\tikzstyle{ave_packet} = [rectangle, draw=GreenOLD, fill=GreenFillD, minimum height=0.6cm, minimum width=1.1cm]

\tikzstyle{large_packet} = [rectangle, draw=GreenOLD, fill=GreenFillD, minimum height=0.6cm, minimum width=1.2cm]

\tikzstyle{xlarge_packet} = [rectangle, draw=GreenOLD, fill=GreenFillD, minimum height=0.6cm, minimum width=1.3cm]

\tikzstyle{xxlarge_packet} = [rectangle, draw=GreenOLD, fill=GreenFillD, minimum height=0.6cm, minimum width=1.4cm]

\tikzstyle{xxxlarge_packet} = [rectangle, draw=GreenOLD, fill=GreenFillD, minimum height=0.6cm, minimum width=1.5cm]

\tikzstyle{max_packet} = [rectangle, draw=GreenOLD, fill=GreenFillD, minimum height=0.6cm, minimum width=1.55cm]

\tikzstyle{inf_packet} = [rectangle, draw=GreenOLD, fill=GreenFillD, minimum height=0.6cm, minimum width=1.8cm]

\tikzstyle{ipg_priority} = [rectangle, draw=PriorityOL, fill=PriorityFill, minimum height=0.5cm, minimum width=0.1cm]

\tikzstyle{ipg1} = [rectangle, draw=OrangeOLD, fill=OrangeFillD, minimum height=0.6cm, minimum width=0.82cm]

\tikzstyle{ipg2} = [rectangle, draw=OrangeOLD, fill=OrangeFillD, minimum height=0.6cm, minimum width=1.22cm]

\tikzstyle{ipg3} = [rectangle, draw=OrangeOLD, fill=OrangeFillD, minimum height=0.6cm, minimum width=1.22cm]

\tikzstyle{priority} = [rectangle, draw=PriorityOL, fill=PriorityFill, minimum height=0.6cm, minimum width=0.5cm]

\tikzstyle{priority2} = [rectangle, draw=PriorityOL, fill=PriorityFill, minimum height=0.6cm, minimum width=0.4cm]

\tikzstyle{collision} = [rectangle, draw=GrayBlueOLD, fill=GrayBlueFillD, minimum height=0.6cm, minimum width=0.2cm]

\tikzstyle{pkt} = [rectangle, draw=GreenOLD, fill=GreenFillD, minimum height=0.45cm, minimum width=0.8cm]

\tikzstyle{beaconStd} = [rectangle, draw=YellowOLD, fill=YellowFillD, minimum height=0.6cm, minimum width=0.4cm]

\tikzstyle{commitStd} = [rectangle, draw=BlueOLD, fill=BlueFillD, minimum height=0.6cm, minimum width=0.4cm]

\tikzstyle{silentStd} = [rectangle, draw=WhiteOLD, fill=WhiteFillL, minimum height=0.6cm, minimum width=0.4cm]

\tikzstyle{dataStd} = [rectangle, draw=GreenOLD, fill=GreenFillD, minimum   height=0.6cm, minimum width=0.4cm]

\tikzstyle{ipgStd} = [rectangle, draw=OrangeOLD, fill=OrangeFillD, minimum height=0.6cm, minimum width=0.4cm]

\tikzstyle{ipg_silentStd} = [rectangle, draw=GrayOLD, fill=GrayFillD, minimum height=0.6cm, minimum width=0.4cm]

\tikzstyle{ipg_priorityStd} = [rectangle, draw=PriorityOL, fill=PriorityFill, minimum height=0.6cm, minimum width=0.4cm]

\begin{figure}[!t]
    \centering    
       \subfloat[Transmission sequence and worst-case signaling overhead in PLCA.]{
        \begin{tikzpicture}

        \begin{scope}[shift={(0, 2.25)}]  
        
        \node () [beaconStd] at (0.5,0) {};
        \node[right] at (0.65, 0) {\tiny BEACON};

        \node () [commitStd] at (2.5,0) {\tiny $n$};
        \node[right] at (2.65, 0) {\tiny COMMIT};

        \node () [silentStd] at (4.5,0) {\tiny $n$};
        \node[right] at (4.65, 0) {\tiny Silent};

        \node () [dataStd] at (6.5,0) {\tiny $n$};
        \node[right] at (6.65, 0) {\tiny Data};
        \end{scope}

        \draw[ ->] (0,0) -- (8.3,0) node[right]{};
        
        \node at (8.1, -0.15) {\tiny Time}; 

        \draw[decorate, decoration={brace, amplitude=1ex, raise=-1ex}] (1.7,0.95)  -- ++ (1,0);
        \node [] at (1.7 + 0.5, 0.95 + 0.15) {\tiny Idle TOs};

        \draw[->] (4.9,1.2) -- (4.9,0.8); 
        \draw[->] (7.1,1.2) -- (7.1,0.8);
        \draw[<->] (4.9 + 0.1,1) -- (7.1 - 0.1,1);
        \node[] at (4.9, 1.2 + 0.2 + 0.2) {\tiny Packet};
        \node[] at (4.9, 1.2 + 0.2) {\tiny Ready};
        \node[] at (4.9 + 1.1, 1 + 0.2) {\tiny $O_{n\mathrm{,}\max}^{\text{PLCA}} $};
        \node[] at (7.1, 1.2 + 0.2 + 0.2) {\tiny Transmit};
        \node[] at (7.1, 1.2 + 0.2) {\tiny Start};

        \node (b1) [beacon] at (0.2,0.4) {};
        \node (c10) [commit, right=0cm of b1] {\tiny 0};
        \node (p10) [small_packet, right=0cm of c10] {\tiny 0};
        \node (s11) [silent, right=0cm of p10] {\tiny 1};
        \node (s12) [silent, right=0cm of s11] {\tiny 2};
        \node (c13) [commit, right=0cm of s12] {\tiny 3};
        \node (p13) [small_packet, right=0cm of c13] {\tiny 3};

        \node (b2) [beacon, right=0cm of p13]{};
        \node (s20) [silent, right=0cm of b2] {\tiny 0};
        \node (s21) [silent, right=0cm of s20] {\tiny 1};
        \node (s22) [silent, right=0cm of s21] {\tiny 2};
        \node (s23) [silent, right=0cm of s22] {\tiny 3};

        \node (b3) [beacon, right=0cm of s23]{};
        \node (c30) [commit, right=0cm of b3] {\tiny 0};
        \node (p30) [small_packet, right=0cm of c30] {\tiny 0};

        \end{tikzpicture}
        \label{subfig: Transmission sequence of standard PLCA}
        }
        \vspace{-3mm}

            \subfloat[Worst-case access latency in PLCA.]{%
        \begin{tikzpicture}

        \draw[ ->] (0,0) -- (8.3,0) node[right]{};
        
        \node at (8.1, -0.15) {\tiny Time};

        \node (b1) [beacon] at (0.2,0.4) {};
        \node (s10) [silent, right=0cm of b1] {\tiny 0};
        \node (c11) [commit, right=0cm of s10] {\tiny 1};
        \node (p11) [max_packet, right=0cm of c11] {\tiny 1};
        \node (c12) [commit, right=0cm of p11] {\tiny 2};
        \node (p12) [max_packet, right=0cm of c12] {\tiny 2};
        \node (c13) [commit, right=0cm of p12] {\tiny 3};
        \node (p13) [max_packet, right=0cm of c13] {\tiny 3};
        
        \node (b2) [beacon, right=0cm of p13]{};
        \node (c20) [commit, right=0cm of b2] {\tiny 0};
        \node (p20) [small_packet, right=0cm of c20] {\tiny 0};

        \draw[->] (0.865,1.2) -- (0.865,0.8); 
        \draw[->] (7.225,1.2) -- (7.225,0.8);
        \draw[<->] (0.865 + 0.1, 1) -- (7.225 - 0.1, 1);
        \node[] at (0.865, 1.2 + 0.2 + 0.2) {\tiny Packet};
        \node[] at (0.865, 1.2 + 0.2) {\tiny Ready};
        \node[] at (0.865 + 3.18, 1 + 0.2) {\tiny $L_{n\mathrm{,}\max}^{\text{PLCA}} $};
        \node[] at (7.225, 1.2 + 0.2 + 0.2) {\tiny Transmit};
        \node[] at (7.225, 1.2 + 0.2) {\tiny Start};

        \end{tikzpicture}
        \label{subfig: Worst-case access latency in PLCA}
    }

    \caption{Operation, worst-case signaling overhead, and worst-case access latency in standard PLCA for a network with $N = 4$ nodes.}
    \label{fig: Operation and transmission sequence of PLCA}
    \vspace{-3mm}

\end{figure}
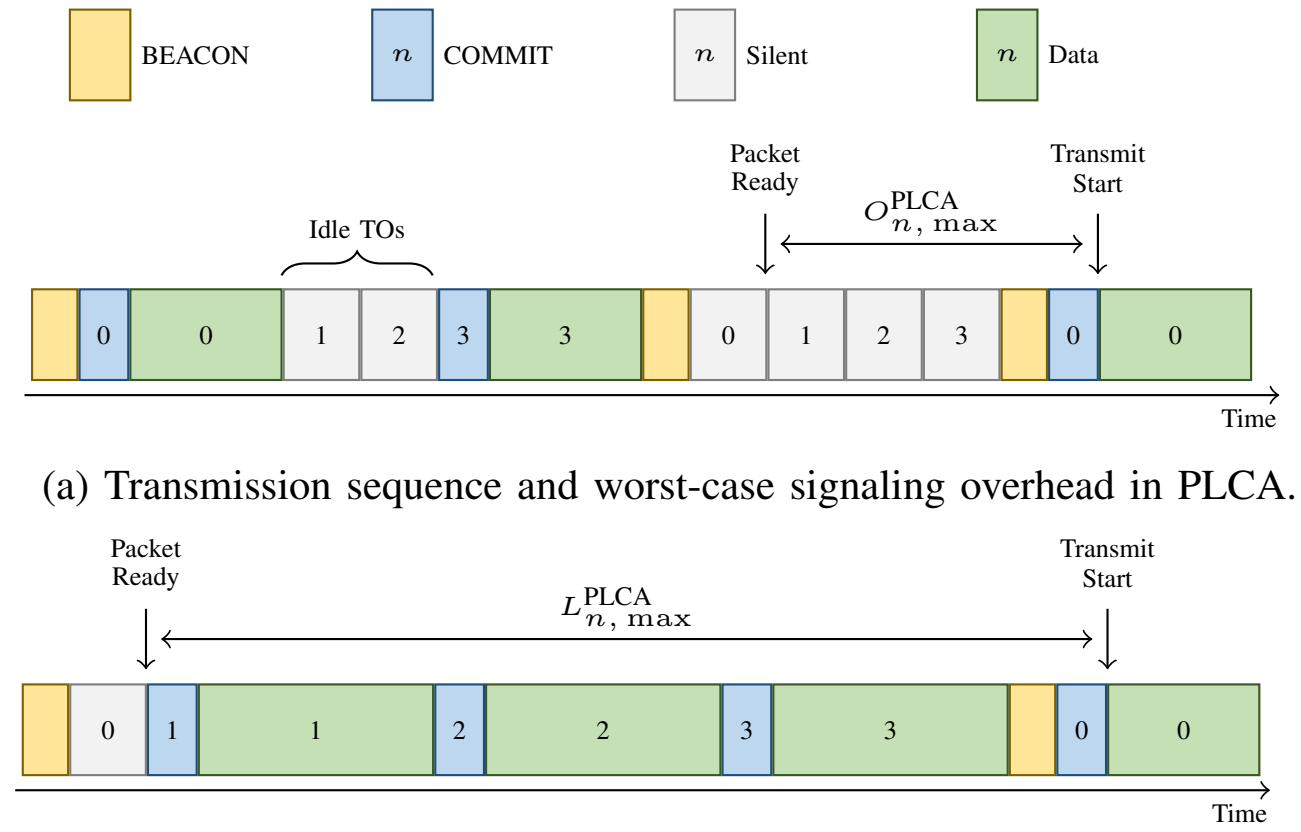

\section{OC-PLCA: Orthogonal Code-Based Parallel Signaling in PLCA}
\label{Section: Novel}

This section introduces OC-PLCA, an orthogonal code-based parallel signaling mechanism for PLCA. It replaces the sequential COMMIT signaling of standard PLCA with a simultaneous COMMIT signaling phase, in which nodes signal their intent to transmit using WH codes. Unlike standard PLCA, the transmission cycle in OC-PLCA includes only the active nodes, eliminating idle TOs entirely. It is important to note that WH codes are used exclusively during the COMMIT signaling phase. Once a node reserves the channel, data transmission proceeds exactly as in standard PLCA.

\definecolor{GrayBlueOLL}{HTML}{44546A}
\definecolor{GrayBlueOLM}{HTML}{333f50}
\definecolor{GrayBlueOLD}{HTML}{222A35}

\definecolor{GrayBlueFillL}{HTML}{D6DCE5}
\definecolor{GrayBlueFillD}{HTML}{ADB9CA}

\definecolor{BlueOLL}{HTML}{5B9BD5}
\definecolor{BlueOLM}{HTML}{2E75B6}
\definecolor{BlueOLD}{HTML}{1F4E79}

\definecolor{BlueFillL}{HTML}{DEEBF7}
\definecolor{BlueFillD}{HTML}{BDD7EE}

\definecolor{GreenOLL}{HTML}{70AD47}
\definecolor{GreenOLM}{HTML}{548235}
\definecolor{GreenOLD}{HTML}{385723}

\definecolor{GreenFillL}{HTML}{E2F0D9}
\definecolor{GreenFillD}{HTML}{C5E0B4}

\definecolor{WhiteOLL}{HTML}{BFBFBF}
\definecolor{WhiteOLM}{HTML}{A6A6A6}
\definecolor{WhiteOLD}{HTML}{7F7F7F}

\definecolor{WhiteFillL}{HTML}{F2F2F2}
\definecolor{WhiteFillD}{HTML}{D9D9D9}

\definecolor{GrayOLL}{HTML}{A5A5A5}
\definecolor{GrayOLM}{HTML}{7C7C7C}
\definecolor{GrayOLD}{HTML}{525252}

\definecolor{GrayFillL}{HTML}{EDEDED}
\definecolor{GrayFillD}{HTML}{DBDBDB}

\definecolor{YellowOLL}{HTML}{FFC000}
\definecolor{YellowOLM}{HTML}{BF9000}
\definecolor{YellowOLD}{HTML}{7F6000}

\definecolor{YellowFillL}{HTML}{FFF2CC}
\definecolor{YellowFillD}{HTML}{FFE699}

\definecolor{OrangeOLL}{HTML}{ED7D31}
\definecolor{OrangeOLM}{HTML}{C55A11}
\definecolor{OrangeOLD}{HTML}{843C0C}

\definecolor{OrangeFillL}{HTML}{FBE5D6}
\definecolor{OrangeFillD}{HTML}{F8CBAD}

\definecolor{PriorityOL}{HTML}{C0392B} 
\definecolor{PriorityFill}{HTML}{F5B7B1}

\definecolor{CollisionOL}{HTML}{16A085} 
\definecolor{CollisionFill}{HTML}{A3E4D7}

\begin{figure}[!t]
    \centering    
       \subfloat[Transmission sequence and worst-case signaling overhead in OC-PLCA.]{
        \begin{tikzpicture}

        \begin{scope}[shift={(0, 2.25)}] 
        
        \node () [beaconStd] at (0.25+0.25,0) {};
        \node[right] at (0.25+0.15+0.25, 0) {\tiny BEACON};

        \node () [commitStd] at (2.5,0) {\tiny $n$};
        \node[right] at (2.5+0.15, 0) {\tiny COMMIT};
        
        \node () [ipg_priorityStd] at (4.5,0) {};
        \node[right] at (4.5+0.15, 0) {\tiny COMMIT Phase};

        \node () [dataStd] at (7+0.25,0) {\tiny $n$};
        \node[right] at (7+0.15+0.25, 0) {\tiny Data};
        \end{scope}

        \draw[ ->] (0,0) -- (8.3,0) node[right]{};
       
        \node at (8.1, -0.15) {\tiny Time}; 

        \node (b1) [beacon] at (0.2,0.4) {};        
        \node(pp1) [priority, right=0cm of b1] {};
        \node (p10) [small_packet, right=0cm of pp1] {\tiny 0};
        \node (p13) [small_packet, right=0cm of p10] {\tiny 3};

        \draw[->] (3.725,1.2) -- (3.725,0.8); 
        \draw[->] (4.55,1.2) -- (4.55,0.8);
        \draw[<->] (3.725 + 0.1, 1) -- (4.55 - 0.1,1);

        \node[] at (3.725, 1.2 + 0.2 + 0.2) {\tiny Packet};
        \node[] at (3.725, 1.2 + 0.2) {\tiny Ready};

        \draw[<-] (3.725 + 0.4125 - 0.02, 0.85 + 0.225 + 0.05) to [bend left=25] (3.725 + 0.825 + 0.825 + 0.825, 0.85+0.225 + 0.05);
        \node[] at (3.725 + 0.825 + 0.825+  0.4125 + 0.4125, 0.85+0.225 - 0.2) {\tiny $O_{n\mathrm{,}\max}^{\text{OC-PLCA}} $};

        \node[] at (4.55, 1.2 + 0.2 + 0.2) {\tiny Transmit};
        \node[] at (4.55, 1.2 + 0.2) {\tiny Start};

        \node[above=-0.2cm of pp1] (card10) [CTO_commit] {\tiny 0};
        \node[above=0.02cm of pp1] (card13) [CTO_commit, xshift=0.2cm] {\tiny 3};

        \node (b2) [beacon, right=0cm of p13]{};
        \node(pp2) [priority, right=0cm of b2] {};

        \node (b3) [beacon, right=0cm of pp2]{};
        \node(pp3) [priority, right=0cm of b3] {};
        \node (p13) [small_packet, right=0cm of pp3] {\tiny 0};
        \node[above=-0.2cm of pp3] (card23) [CTO_commit] {\tiny 0};

        \end{tikzpicture}
        \label{subfig: Transmission sequence of OC-PLCA}
        }
        \vspace{-3mm}
            \subfloat[Worst-case access latency in OC-PLCA.]{%
        \begin{tikzpicture}

        \draw[ ->] (0,0) -- (8.3,0) node[right]{};
        \node at (8.1, -0.15) {\tiny Time};

        \node (b1) [beacon] at (0.2,0.4) {};
        \node(pp1) [priority, right=0cm of b1] {};
        
        \node (c12) [max_packet, right=0cm of pp1] {\tiny 1};
        \node (p12) [max_packet, right=0cm of c12] {\tiny 2};
        \node (p13) [max_packet, right=0cm of p12] {\tiny 3};

        \node (b2) [beacon, right=0cm of p13]{};
        \node(pp2) [priority, right=0cm of b2] {};

        \node (c20) [small_packet, right=0cm of pp2] {\tiny 0};

        \draw[->] (0.875,1.2) -- (0.875,0.8); 
        \draw[->] (6.4,1.2) -- (6.4,0.8);
        \draw[<->] (0.875 + 0.1, 1) -- (6.4 - 0.1, 1);
        \node[] at (0.875, 1.2 + 0.2 + 0.2) {\tiny Packet};
        \node[] at (0.875, 1.2 + 0.2) {\tiny Ready};
        \node[] at (0.875 + 2.7625, 1 + 0.2) {\tiny $L_{n\mathrm{,}\max}^{\text{OC-PLCA}} $};
        \node[] at (6.4, 1.2 + 0.2 + 0.2) {\tiny Transmit};
        \node[] at (6.4, 1.2 + 0.2) {\tiny Start};

        \node[above=-0.25cm of pp1] (card13) [CTO_commit, xshift=+0.2cm] {\tiny 3};
        \node[above=-0.01cm of pp1] (card12) [CTO_commit] {\tiny 2};
        \node[above=-0.25cm of pp1] (card11) [CTO_commit, xshift=-0.2cm] {\tiny 1};

        \node[above=-0.2cm of pp2] (card20) [CTO_commit] {\tiny 0};

        \end{tikzpicture}
        \label{subfig: Worst-case access latency in OC-PLCA}
    }
    
    \caption{Operation, worst-case signaling overhead, and worst-case access latency in OC-PLCA for a network with $N = 4$ nodes.}
    \label{fig: Operation and transmission sequence of OC-PLCA}
    \vspace{-3mm}
\end{figure}
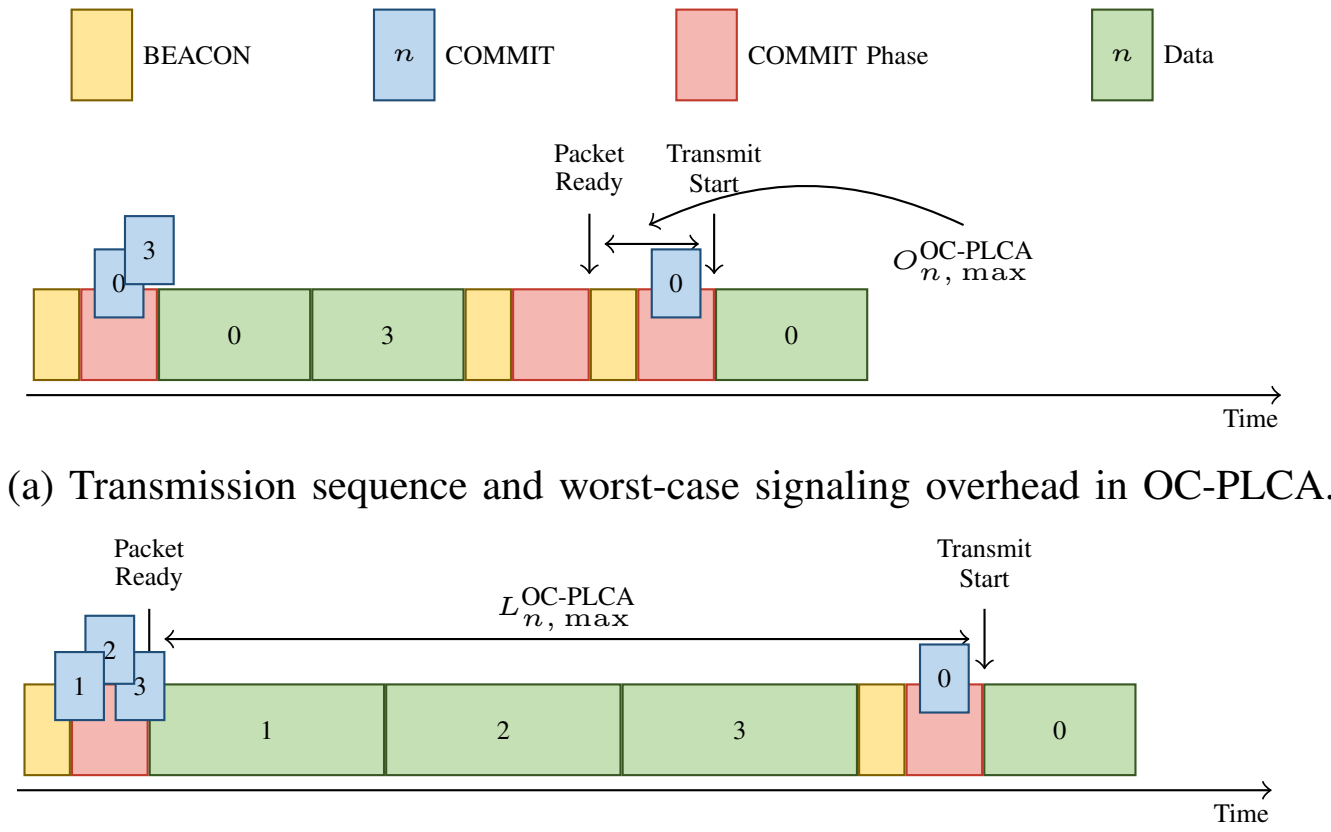

The code assignment scheme is first presented, followed by the OC-PLCA operation and cycle structure. The access latency characteristics are then derived. Next, active node detection under quantization is investigated. Finally, a grouping-based extension to support large networks is proposed.

\subsection{Code Assignment}
\label{Subsection: Code Assignment}

Each node $n \in \mathcal{N}$ is assigned a unique codeword from~$\mathcal{C}$, which it uses exclusively during the COMMIT signaling phase to signal its intent to transmit. Since the codebook size is constrained to powers of two, a codebook of size $2^m$, where $m = \lceil \log_2 N \rceil$, is required to support a network of $N$ nodes.

While an arbitrary assignment of codewords to nodes would preserve the orthogonality property, we propose a deterministic assignment method based on zero-crossing ordering\footnote{The zero-crossing count of a codeword is defined as the number of transitions between $+1$ and $-1$ in the code sequence.} for distributed operation and backward compatibility. Node~$n$ is assigned the codeword with $n$ zero crossings. Henceforth, we denote the codeword assigned to node $n$ as $\mathbf{c}_n$, where $n$ reflects both the node ID and the zero-crossing count of the assigned codeword. For example, for $m=2$, the assignment~is:
\begin{equation}
\begin{array}{ll}
\mathbf{c}_0 = [+1,+1,+1,+1]^T \rightarrow \text{node } n= 0 \; (0 \text{ crossings})\\
\mathbf{c}_1 = [+1,+1,-1,-1]^T \rightarrow \text{node } n= 1 \; (1 \text{ crossing})\\
\mathbf{c}_2 = [+1,-1,-1,+1]^T \rightarrow \text{node } n= 2 \; (2 \text{ crossings})\\
\mathbf{c}_3 = [+1,-1,+1,-1]^T \rightarrow \text{node } n=3 \; (3 \text{ crossings})
\end{array}
\end{equation}

\subsection{OC-PLCA Operation}
\label{Subsection: OC-PLCA Operation}

An access cycle begins with a BEACON message, synchronizing all devices, as in standard PLCA. A parallel COMMIT phase is then initiated, during which nodes can simultaneously signal their intent to transmit. Nodes with pending traffic transmit their assigned codeword $\mathbf{c}_n$ of duration~$T_\mathrm{p}$:
\begin{equation}
T_\mathrm{p} = 2^m\cdot T_\mathrm{s},
\label{equ: basic Tp}
\end{equation}
while nodes with no pending traffic remain silent. If the phase remains entirely silent, the cycle is terminated after the predefined idle interval~$T_\mathrm{i}$.

Following the parallel COMMIT phase, each node independently identifies the active set $\mathcal{A}$ by correlating the received signal $\mathbf{r}$ with each codeword. A node~$n$ is identified as active if $y_n= 2^m$, and inactive otherwise.

Since active-node identification occurs once at the beginning of the cycle, a node may be granted two successive TOs while another active node remains unserved. To ensure fairness, the current cycle's order $\pi$ is constructed from the current cycle's active set~$\mathcal{A}$ and the previous cycle's active set~$\mathcal{A}^{-}$ and order $\pi^{-}$. The nodes in $\mathcal{A} \backslash \mathcal{A}^{-}$ (i.e., active nodes that did not transmit in the previous cycle) are listed first, in ascending order of node ID, followed by the nodes in $\mathcal{A} \cap \mathcal{A}^{-}$ (i.e., active nodes that did transmit in the previous cycle), retaining their relative order from $\pi^{-}$.

Upon constructing $\pi$, each node initializes a local counter $\hat{a}=0$, tracking the position of the current TO within the cycle. When $\hat{a}$ matches a node's position in $\pi$, that node transmits its pending packet, after which $\hat{a}$ is incremented, advancing to the next node. Once the last TO is completed ($\hat{a} = |\mathcal{A}|-1$), the cycle ends, $\pi^{-} \leftarrow \pi$, $\mathcal{A}^{-} \leftarrow \mathcal{A}$, and a new BEACON is issued to start the next cycle. The transmission sequence in OC-PLCA is illustrated in Fig.~\ref{fig: Operation and transmission sequence of OC-PLCA}.

\subsection{Latency Analysis}
\label{Subsection: Latency Analysis}

OC-PLCA improves access latency and channel utilization under both heavy and light load conditions, as shown in Fig.~\ref{fig: Operation and transmission sequence of OC-PLCA}. Under heavy load, the worst-case access latency occurs when all $N-1$ preceding nodes are active and transmit maximum-sized packets, which can be expressed as: 
\begin{equation}
\label{equ: OC-PLCA maximum latency}
L_{n,\max}^{\mathrm{OC-PLCA}} = T_\mathrm{b} + T_{\mathrm{p}} + (N-1) T_{\mathrm{pkt,max}},
\end{equation}
replacing the $N$ sequential COMMIT messages of standard PLCA with a single parallel phase. Under light load, the worst-case signaling overhead occurs when all $N-1$ preceding nodes have no pending data, resulting in:
\begin{equation}
\label{equ: OC-PLCA maximum overhead}
O_{n,\max}^{\mathrm{OC-PLCA}} = T_\mathrm{b} + T_\mathrm{p}.
\end{equation}
Compared to (\ref{equ: PLCA maximum overhead}), the $(N-1)$ idle intervals of standard PLCA are eliminated entirely. Since $T_\mathrm{i}$ is significantly larger than~$T_\mathrm{c}$, as shown in Table~\ref{tab:parameters}, this elimination yields a substantially larger reduction in signaling overhead than that achieved under heavy load.

\begin{table}[t]
\vspace{-3mm}
\centering
\renewcommand{\arraystretch}{1}
\caption{Standard 10BASE-T1S Parameter Values \cite{8023cg}.}
\label{tab:parameters}
\begin{tabular}{lccc}
\hline
\textbf{Parameter} & \textbf{Symbol} & \textbf{Size} & \textbf{Duration} \\
\hline
BEACON duration       & $T_\mathrm{b}$ & 20 bits & $2$~$\mu$s \\
COMMIT duration       & $T_\mathrm{c}$ & 5 bits & $0.5$~$\mu$s \\
Idle interval duration & $T_\mathrm{i}$ & 32 bit times & $3.2$~$\mu$s \\
\hline
\end{tabular}
\vspace{-3mm}
\end{table}

\subsection{Quantization Impact on Active Node Detection}
\label{Subsection: PHY Considerations}

This section investigates whether active nodes can be unambiguously identified when the receiver observes only a quantized version of the superimposed codewords. The correlation-based detection described in Sections~\ref{Subsection: Walsh-Hadamard Codes} and~\ref{Subsection: OC-PLCA Operation} assumes that the receiver resolves the received signal $\mathbf{r}$ to its exact integer value, preserving the linear superposition of simultaneously transmitted codewords, a condition satisfied only by a high-precision quantizer. However, the baseband processing of current Single Pair Ethernet (SPE) standards typically employs simpler three-level or two-level quantizers, mapping the received signal to $\{+1,0,-1\}$ or $\{+1,-1\}$, respectively. This low-precision quantization introduces a nonlinear mapping that degrades the correlator output, potentially producing false active or false inactive detections.

Despite this nonlinearity, we demonstrate in the following that active nodes can still be effectively identified. The key requirement for correct identification is that the mapping from active node set to quantized received signal be injective, i.e., distinct active node sets produce distinct quantized outputs. If this condition holds, active nodes can be correctly identified, for instance, through runtime correlation or an offline lookup table (LUT).

Let $\mathbf{b}=[b_0, b_1, \dots, b_{N-1}]^T \in \{0,1\}^N$ denote the binary state vector, where $b_n = 1$ if node $n$ is active and $b_n = 0$ otherwise. Assuming unity channel gain and negligible noise for simplicity, the received signal in (\ref{equ: received signal r}) can be equivalently expressed as:
\begin{equation}
\mathbf{r} = \sum_{n=0}^{N-1}{b_n \mathbf{c}_n}=\mathbf{C}\mathbf{b},
\end{equation}
where $\mathbf{C}=[\mathbf{c}_0, \mathbf{c}_1, \dots , \mathbf{c}_{N-1}]$ denotes the matrix whose columns are the codewords assigned to each node. The quantized received signal $\tilde{\mathbf{r}}$ is given by:
\begin{equation}
\tilde{\mathbf{r}} =f(\mathbf{b}) = Q(\mathbf{C}\mathbf{b}),
\end{equation}
where $Q(\cdot)$ denotes the scalar quantizer under consideration, applied component-wise to $\mathbf{C}\mathbf{b}$. For an observed vector $\tilde{\mathbf{r}}$, we define its ambiguity order as:
\begin{equation}
\alpha(\tilde{\mathbf{r}}) = |\{\mathbf{b} \in \{0,1\}^N : f(\mathbf{b}) = \tilde{\mathbf{r}}\}|,
\end{equation}
representing the number of state vectors $\mathbf{b}$ that produce the same $\tilde{\mathbf{r}}$. The active node state is uniquely identified if and only if the forward mapping $\mathbf{b} \to \tilde{\mathbf{r}}$ is injective, equivalently:
\begin{equation}
\alpha(\tilde{\mathbf{r}}) = 1, \qquad \forall\, \tilde{\mathbf{r}} \in \text{Im}(f).
\label{eq:injectivity_condition}
\end{equation}
If this condition holds, the inverse mapping $f^{-1}$ is well-defined over $\text{Im}(f)$ and can be implemented by a LUT, using a ternary key:
\begin{equation}
K_{\tilde{r}} = \sum_{k=0}^{2^m-1} (\tilde{r}_k + 1) \cdot 3^k,
\end{equation}
and a binary key:
\begin{equation}
K_b = \sum_{n=0}^{N-1} b_n \cdot 2^n.
\end{equation}

\subsubsection{Three-Level Quantizer}
\label{Subsubsection: Three-Level Decoder}

Under the three-level quantizer, each element $r_k$ of the received signal $\mathbf{r}$ is quantized according to a squelch threshold $\tau$ as:
\begin{equation}
    \widetilde{r}_k
    =
    \begin{cases}
        +1, & r_k>\tau,\\
        -1, & r_k<-\tau,\\
        0,  & \text{otherwise,}
    \end{cases}
    \qquad 0<\tau<1 .
    \label{eq:three_level_quantizer}
\end{equation}
Table~\ref{tab:LUT_3_2_level_quantizers} shows the resulting LUT for $m=2$ (i.e., 4 codewords). Since all observed ternary keys $K_{\tilde{r}}$ are distinct, the injectivity condition holds, and active nodes can be unambiguously identified.

Exhaustive enumeration confirms injectivity for all codebooks up to $\mathbf{H}_3$ (i.e., $2^3=8$ codewords). For $m>3$, however, injectivity does not hold. Thus, to preserve injectivity for arbitrary~$N$, a grouping-based extension, introduced in Section~\ref{Subsection: Grouping}, partitions the network into groups of at most eight nodes served during independent phases using $\mathbf{H}_3$.

\begin{table}[t]
\vspace{-3mm}
\centering
\setlength{\tabcolsep}{2.5pt}
\caption{LUT for the three-level quantizer ($m=2$, $\tau=0.5$) and two-level quantizer ($m=2$).}
\label{tab:LUT_3_2_level_quantizers}
\begin{tabular}{ccccccc}
\hline
& & & \multicolumn{2}{c}{\textbf{Three-level}} 
  & \multicolumn{2}{c}{\textbf{Two-level}} \\

$K_b$ & $\mathbf{b}$ & $\mathbf{r}$
& $\tilde{\mathbf{r}}$ & $K_{\tilde{\mathbf{r}}}$
& $\tilde{\mathbf{r}}$ & $K_{\tilde{\mathbf{r}}}$ \\

\hline

0 & $[0,0,0,0]$ & $[0,0,0,0]$
& $[0,0,0,0]$ & $40$
& $[+,+,+,+]$ & $80$ \\

1 & $[1,0,0,0]$ & $[+1,+1,+1,+1]$
& $[+,+,+,+]$ & $80$
& \cellcolor{gray!10}$[+,+,+,+]$ & \cellcolor{gray!10}$80$ \\

2 & $[0,1,0,0]$ & $[+1,+1,-1,-1]$
& $[+,+,-,-]$ & $8$
& $[+,+,-,-]$ & $8$ \\

3 & $[1,1,0,0]$ & $[+2,+2,0,0]$
& $[+,+,0,0]$ & $44$
& \cellcolor{gray!10}$[+,+,+,+]$ & \cellcolor{gray!10}$80$ \\

4 & $[0,0,1,0]$ & $[+1,-1,-1,+1]$
& $[+,-,-,+]$ & $56$
& $[+,-,-,+]$ & $56$ \\

5 & $[1,0,1,0]$ & $[+2,0,0,+2]$
& $[+,0,0,+]$ & $68$
& \cellcolor{gray!10}$[+,+,+,+]$ & \cellcolor{gray!10}$80$ \\

6 & $[0,1,1,0]$ & $[+2,0,-2,0]$
& $[+,0,-,0]$ & $32$
& $[+,+,-,+]$ & $62$ \\

7 & $[1,1,1,0]$ & $[+3,+1,-1,+1]$
& $[+,+,-,+]$ & $62$
& \cellcolor{gray!10}$[+,+,-,+]$ & \cellcolor{gray!10}$62$ \\

8 & $[0,0,0,1]$ & $[+1,-1,+1,-1]$
& $[+,-,+,-]$ & $20$
& $[+,-,+,-]$ & $20$ \\

9 & $[1,0,0,1]$ & $[+2,0,+2,0]$
& $[+,0,+,0]$ & $50$
& \cellcolor{gray!10}$[+,+,+,+]$ & \cellcolor{gray!10}$80$ \\

10 & $[0,1,0,1]$ & $[+2,0,0,-2]$
& $[+,0,0,-]$ & $14$
& $[+,+,+,-]$ & $26$ \\

11 & $[1,1,0,1]$ & $[+3,+1,+1,-1]$
& $[+,+,+,-]$ & $26$
& \cellcolor{gray!10}$[+,+,+,-]$ & \cellcolor{gray!10}$26$ \\

12 & $[0,0,1,1]$ & $[+2,-2,0,0]$
& $[+,-,0,0]$ & $38$
& $[+,-,+,+]$ & $74$ \\

13 & $[1,0,1,1]$ & $[+3,-1,+1,+1]$
& $[+,-,+,+]$ & $74$
& \cellcolor{gray!10}$[+,-,+,+]$ & \cellcolor{gray!10}$74$ \\

14 & $[0,1,1,1]$ & $[+3,-1,-1,-1]$
& $[+,-,-,-]$ & $2$
& $[+,-,-,-]$ & $2$ \\

15 & $[1,1,1,1]$ & $[+4,0,0,0]$
& $[+,0,0,0]$ & $41$
& \cellcolor{gray!10}$[+,+,+,+]$ & \cellcolor{gray!10}$80$ \\

\hline
\end{tabular}
\begin{tablenotes}
\footnotesize
\item $+$ and $-$ denote $+1$ and $-1$, respectively.
\end{tablenotes}
\vspace{-3mm}
\end{table}

\subsubsection{Two-Level Quantizer}
\label{Subsubsection: Two-Level Decoder}
Under the two-level quantizer, each element $r_k$ of the received signal $\mathbf{r}$ is quantized as:
\begin{equation}
\tilde{r}_k = \begin{cases} +1, & r_k \geq 0, \\ -1, & r_k < 0. \end{cases}
\label{equ: two-level quantization}
\end{equation}
Since this quantization does not distinguish a zero result from a positive one, the codeword $\mathbf{c}_0 = [+1, +1, \dots, +1]^T$ is indistinguishable from the inactive state:
\begin{equation}
Q(\mathbf{0}_{2^m}) = Q(\mathbf{c}_0) = \mathbf{1}_{2^m}.
\end{equation}
We therefore exclude $\mathbf{c}_0$ from the codebook, and relabel the remaining codewords from $0$ to $2^m-2$. A codebook with $m = \lceil \log_2(N+1) \rceil$ is thus required to support $N$ nodes.

The ambiguity introduced by $\mathbf{c}_0$ is illustrated in Table~\ref{tab:LUT_3_2_level_quantizers}. Excluding the shaded rows, where $b_0=1$, the remaining entries yield distinct ternary keys, confirming an injective mapping. Exhaustive enumeration further verifies injectivity for all codebooks up to $\mathbf{H}_3$ (i.e., $2^3-1=7$ usable codewords) when $\mathbf{c}_0$ is excluded.

\subsection{Grouping-based Extension}
\label{Subsection: Grouping}

To preserve injectivity for arbitrary $N$, the network nodes~$\mathcal{N}$ are partitioned into $\lceil N/S \rceil$ groups $\mathcal{G}_g$:
\begin{equation}
\mathcal{G}_g = \{n\in\mathcal{N}: \left \lfloor \frac{n}{S} \right \rfloor  = g \},
\end{equation}
where $S$ denotes the maximum number of nodes per group, corresponding to the number of usable codewords under the quantizer in use (i.e., $S=8$ for the three-level quantizer, and $S=7$ for the two-level quantizer). Each group $\mathcal{G}_g$ is served by an independent COMMIT phase $g \in \{0,1,\dots,\lceil N/S \rceil-1\}$, ensuring that no more than $S$ codewords are transmitted simultaneously. Since groups transmit during distinct, non-overlapping phases, the same $\mathbf{H}_3$ codebook can be reused across groups without inter-group interference. Consistent with the zero-crossing-based assignment, node $n \in \mathcal{G}_g$ is assigned codeword $c_{n~\mathrm{mod}~S}$.

The access cycle thus comprises a BEACON message, $\lceil N/S \rceil$ COMMIT phases, and a data transmission phase. During phase $g$, only active nodes in $\mathcal{G}_g$ transmit their assigned codewords, while all remaining nodes observe the channel. Each node independently identifies $\mathcal{A}_g \subseteq \mathcal{G}_g$ from the received vector $\mathbf{r}_g$ observed during phase $g$. After all phases complete, the full active set is given by:
\begin{equation}
\mathcal{A} = \bigcup_{g=0}^{\lceil N/S \rceil - 1} \mathcal{A}_g.
\end{equation}
Active nodes then transmit their data packets according to the cycle order $\pi$, as described in Section~\ref{Subsection: OC-PLCA Operation}.

The worst-case access latency and signaling overhead under the grouping-based extension are given by (\ref{equ: OC-PLCA maximum latency}) and (\ref{equ: OC-PLCA maximum overhead}), respectively, with the total COMMIT phase duration $T_\mathrm{p}$ defined~as:
\begin{equation}
    T_\mathrm{p} = \left \lceil \frac{N}{S} \right \rceil \cdot 2^3 \cdot T_\mathrm{s}.
\label{equ: grouping Tp}
\end{equation}

\section{Evaluation} 
\label{Section: Evaluation}

This section evaluates the performance of OC-PLCA against standard PLCA through theoretical and simulation analysis.

\subsection{Theoretical Performance and Scalability Analysis}
\label{Subsection: Theoretical Analysis}

\begin{figure*}[tp]
    \vspace{-3mm}
    \centering
    \includegraphics[width=\linewidth]{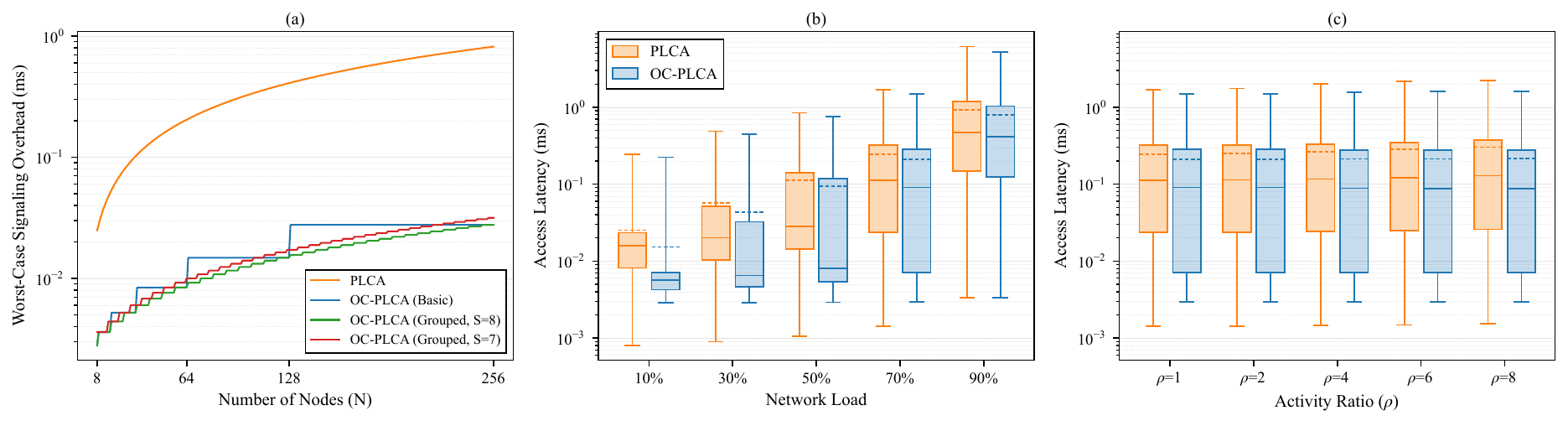}
    \caption{Theoretical and simulation results comparing PLCA and OC-PLCA: (a) worst-case signaling overhead as a function of network size, (b) observed access latency under varying network load, and (c) observed access latency under heterogeneous node activity at $70\%$ network load.}
    \label{fig: Theoretical and simulation results}
    \vspace{-3mm}
\end{figure*}

We first evaluate the worst-case signaling overhead of PLCA, given by (\ref{equ: PLCA maximum overhead}), and of OC-PLCA, given by (\ref{equ: OC-PLCA maximum overhead}), with $T_\mathrm{p}$ as defined in (\ref{equ: basic Tp}) for basic OC-PLCA and in (\ref{equ: grouping Tp}) for the grouping-based extension. Fig.~\ref{fig: Theoretical and simulation results}a shows the resulting worst-case overhead as a function of network size for PLCA, basic OC-PLCA, and the grouping-based extension with $S=8$ (three-level quantizer) and $S=7$ (two-level quantizer).

The results show that OC-PLCA substantially reduces signaling overhead compared to PLCA and scales significantly better with network size. At $N=8$, OC-PLCA reduces the worst-case signaling overhead by a factor of $8.9$ relative to PLCA ($88.76$\% reduction), and this gap widens at $N=256$ to a factor of $29.7$ ($96.63$\% reduction). Comparing basic OC-PLCA with the grouping-based extension, the latter provides finer-grained scaling, stepping by $2^3 T_\mathrm{s}$ every~$S$ nodes, compared to the power-of-two granularity of basic OC-PLCA.

\subsection{Simulation-based Analysis}
\label{Subsection: Simulation-based Analysis}

We next validate the performance advantages of OC-PLCA through simulation using OMNEST~\cite{omnest}. The simulated network consists of $N=8$ nodes connected via a common channel operating at 10~Mbps. Traffic is generated independently at each device according to a Poisson process, with exponential inter-arrival times. Packet sizes follow a truncated exponential distribution with a mean of $150$~B, bounded between $46$~B and $1500$~B. Network load is varied across five levels ($10\%$ to $90\%$) of the channel capacity. For each load, $25$ independent simulation runs are performed, each using a different seed and a duration of $100$~s.

Fig.~\ref{fig: Theoretical and simulation results}b presents the observed access latency of PLCA and OC-PLCA across the five loads, with whiskers indicating the minimum and maximum observed latency, the box covering the 25th to 75th percentiles, the dashed line indicating the mean, and the solid line indicating the median. OC-PLCA substantially reduces observed access latency across all loads, with reductions in mean latency ranging from $14.4\%$ to $39.6\%$, and in maximum latency ranging from $8.8\%$ to $14.5\%$. The reduction is most pronounced at low loads, where idle TOs dominate PLCA's access cycle and are eliminated entirely by OC-PLCA.

We further evaluate OC-PLCA under heterogeneous traffic scenarios with asymmetric node activity. One node is designated as the active node, generating traffic at a rate $\rho$ times that of each of the remaining nodes, with $\rho=\{1,2,4,6,8\}$, at a fixed network load of $70\%$. As shown in Fig.~\ref{fig: Theoretical and simulation results}c, OC-PLCA's advantage over PLCA persists, and even grows, under high network load when node activity is non-uniform. OC-PLCA reduces mean access latency from around $14\%$ at $\rho=1$ (uniform activity) to up to $29.2\%$ under skewed activity, and maximum observed latency from around $12\%$ to up to $28.4\%$.

\section{Conclusion}  
\label{Section: Conclusion}

Although PLCA provides collision-free medium access in shared-medium Ethernet, its sequential, node-based access structure introduces significant idle time and overhead, a limitation inherited by existing PLCA extensions. To address this, we proposed OC-PLCA, where nodes signal their intent to transmit simultaneously using WH codes, allowing the access cycle to include only active nodes and eliminating idle TOs entirely. We further investigated the impact of quantization on active node detection, and proposed a grouping-based extension that preserves injectivity for networks of an arbitrary number of nodes. Both the theoretical and simulation analyses showed that OC-PLCA substantially reduces signaling overhead and access latency compared to PLCA. Future work will address further practical physical-layer effects, including synchronization errors, propagation delays, and other deployment-related effects.

\bibliographystyle{IEEEtran}
\bibliography{IEEEabrv,reference}

\end{document}